\documentclass[12pt,a4paper]{article} 
\usepackage{hyperref}
\usepackage{amssymb,amsmath}
\usepackage{graphicx}
\def\Rey{R}

\def\Reg{\Rey_{\rm g}}
\def\Ret{\Rey_{\rm t}}
\def\Rec{\Rey_{\rm c}}
\def\bnabla{\boldsymbol{\nabla}}

\title{Laminar-Turbulent Patterning in Transitional Flows}

\author{{\Large Paul Manneville}\\[1ex]
Hydrodynamics Laboratory,\\
CNRS UMR 7646, \'Ecole Polytechnique, Palaiseau, France\\[0.5ex]
{\tt paul.manneville@ladhyx.polytechnique.fr}}
\date{{\sl Entropy} {\bf2017}, {\it19}, 316; doi:10.3390/e19070316 {\normalsize (open access)}\\[2ex]
\normalsize Special Issue\\[0.5ex]
 ``Complex Systems, Non-Equilibrium Dynamics and Self-Organisation.''\\
{\normalsize Guest Editor: Dr. Gunnar Pr\"ussner}\\[1ex]
\today}
\begin{document}
\maketitle
\sloppy

\begin{abstract}
\noindent Wall-bounded flows experience a transition to turbulence characterized by the coexistence of laminar and turbulent domains in some range of Reynolds number $\Rey$, the natural control parameter.
This transitional regime takes place between an upper threshold $\Ret$ above which turbulence is uniform (featureless) and a lower threshold $\Reg$ below which any form of turbulence decays, possibly at the end of overlong chaotic transients.
The most emblematic cases of flow along flat plates transiting to/from turbulence according to this scenario are reviewed.
The coexistence is generally in the form of bands, alternatively laminar and turbulent, and oriented obliquely with respect to the general flow direction.
The final decay of the bands at $\Reg$ points to the relevance of directed percolation and criticality in the sense of statistical-physics phase transitions.
The nature of the transition at $\Ret$ where bands form is still somewhat mysterious and does not easily fit the scheme holding for pattern-forming instabilities at increasing control parameter on a laminar background.
In contrast, the bands arise at $\Ret$ out of a uniform turbulent background at a decreasing control parameter.
Ingredients of a possible theory of laminar-turbulent patterning are discussed.\\[2ex]
\noindent{\bf keywords}: transition to/from turbulence; wall-bounded shear flow; plane Couette flow; turbulent patterning; phase transitions; directed percolation
\end{abstract}

\vspace{2ex}

\noindent The present Special-Issue contribution deals with the transition to turbulence in wall-bounded flows, an important case of systems driven far from equilibrium where patterns develop against a turbulent background.
This active field of research is rapidly evolving, and important results have been obtained recently.
To set the frame, in Section \ref{S1}, I will summarize a recent paper reviewing the subject from a more general standpoint \cite{Ma16a}, enabling me to focus on a specific feature of this transition: the existence of a statistically well-organized laminar-turbulent patterning of flows along planar walls in some intermediate range of Reynolds numbers $[\Reg,\Ret]$. 
The Reynolds number is the main {\emph{control parameter}} of the problem. Its generic expression reads $\Rey=V \ell/\nu$, in which $V$ and $\ell$ are typical velocity and length scales, and $\nu$ the fluid's kinematic viscosity. $\Rey$ compares the typical shear rate $V/\ell$ to the viscous diffusion rate over the same length scale $\nu/\ell^2$.
 $\Reg$  is a global stability threshold marking unconditional return to laminar flow and $\Ret$ some upper threshold beyond which turbulence is essentially uniform.
After having taken the cylindrical shear configuration as an illustrating case in~Section~\ref{S2}, I will turn to strictly planar cases in Section \ref{S3}.
The best understood part of the transition scenario, pattern decay at $\Reg$ is considered in~Section~\ref{S4}.
How patterns emerge as $\Rey$ is decreased from large values is next examined in Section~\ref{S5} before a discussion of perspectives and questions that, in my view, remain open in~Section~\ref{S6}.
I have tried to limit the bibliography to contributions of specific significance, historical or physical, and to the most recent articles of which I am aware.
The remaining plethoric literature on the subject can be accessed via the review articles or books quoted, which also introduce background prerequisites when necessary.

\section{Context\label{S1}}
Under weak forcing, close to thermodynamic equilibrium, fluid motion is \emph{laminar}, i.e., smoothly evolving in space and time with macroscopic transfer properties of microscopic origin (molecular dissipation).
When driven sufficiently far from equilibrium, the flow generically becomes \emph{turbulent}, with irregular swirls on a wide continuum of spatiotemporal scales and enormously enhanced effective transport properties.
The full Navier-Stokes system, i.e. the set formed by the equations governing the velocity and pressure fields $\{\mathbf v, p\}$, $\rho(\partial_t{\mathbf v} + {\mathbf v}\cdot\bnabla {\mathbf v}) = -\bnabla p + \eta \bnabla^2 {\mathbf v}$\quad ($\rho\,:$ density; $\eta=\rho\nu\,:$ dynamic viscosity; ${\mathbf v}\cdot\bnabla {\mathbf v}\,:$ advection term) and the continuity equation that simply reads $\bnabla\cdot {\mathbf v} = 0$ for the incompressible flow of simple fluids, plus boundary and initial conditions, called the Navier-Stokes equation (NSE) for short in the following, governs the whole flow behavior.
As the applied shear rate increases, its viscous (Stokes) part is overtaken by its nonlinear advection term that enables nontrivial solutions competing with the unique trivial \emph{base state} permitted near equilibrium.
The transition from laminar to turbulent dynamics has been an important field of study, in view of deep theoretical issues relating to the nature of stochasticity and its important consequences on macroscopic transfer properties in applications (consult \cite{Ma10} for an introduction).
Basically two transition scenarios can be distinguished upon varying $\Rey$ \cite{Ma16a}.

\begin{figure}
\includegraphics[width=\textwidth]{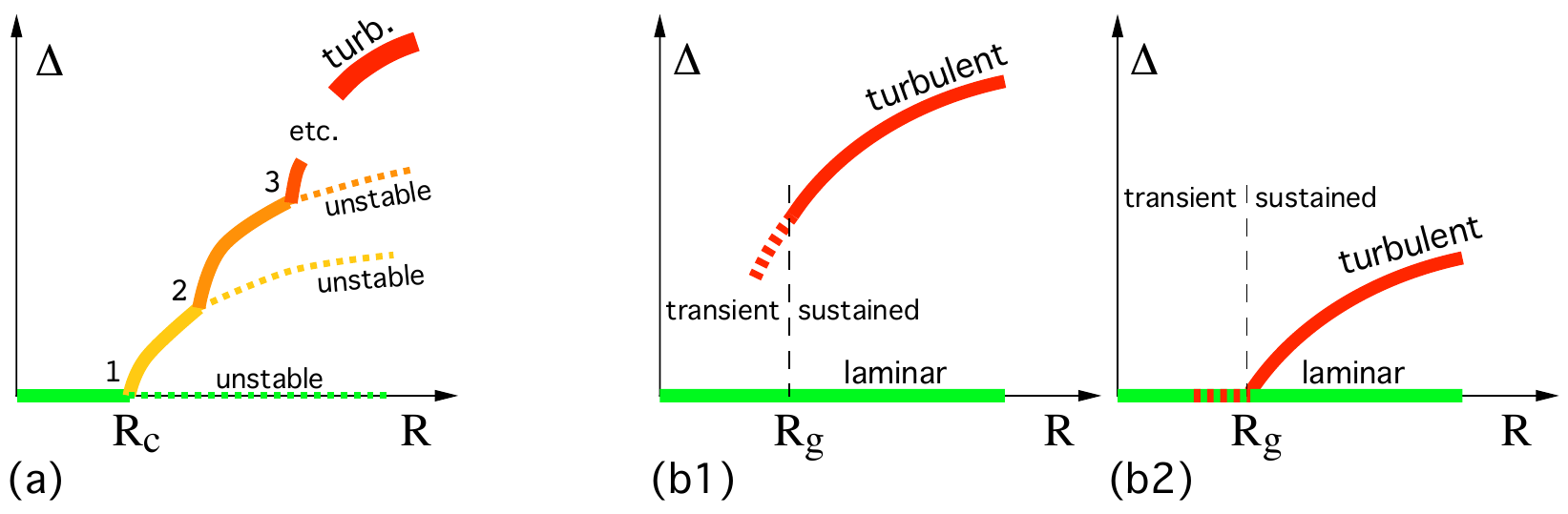}
\caption{(\textbf{a}) Globally super-critical scenario: more and more modes become progressively active before the flow can be considered turbulent; (\textbf{b}) Globally sub-critical scenario. Qualitatively, sufficiently large perturbations are needed to reach the turbulent branch. Quantitatively, a distance $\Delta$ to the laminar branch can be defined, but may vary with $\Rey$ discontinuously (\textbf{b1}) or continuously (\textbf{b2}) depending on whether fully-localized coherent structures are long-lived or not, hence whether the turbulent fraction measured in an infinitely-extended system can tend to zero, Case b2 (to be discussed in Section~\ref{S4}).\label{f1}}
\end{figure}

\begin{figure}
\includegraphics[width=\textwidth]{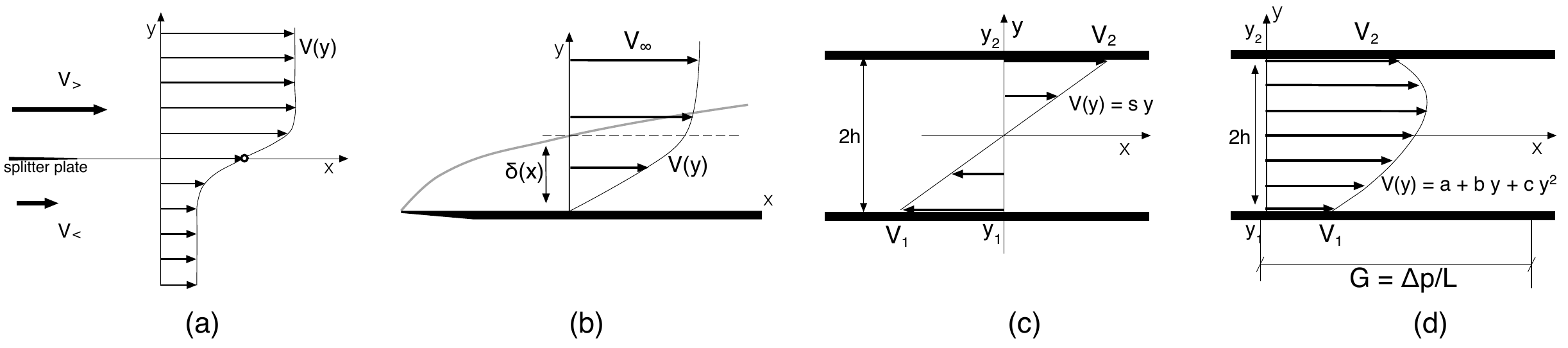}
\caption{(\textbf{a}) Kelvin--Helmholtz instability of an inflectional velocity profile is mostly responsible for laminar breakdown at low $\Rey$, here in a mixing layer down a splitter plate. (\textbf{b}--\textbf{d}) Non-inflectional velocity profiles. ({\textbf{b}}) The Blasius boundary layer velocity profile scales as the square root of the distance to the plate's leading edge. This downstream evolution can be suppressed by suction through the plate when porous. ({\textbf{c}}) The plane Couette flow displays a linear velocity profile, with shearing rate $s=(V_2-V_1)/2h$; here $V_1=-V_2$, hence no mean advection. ({\textbf{d}}) The Couette--Poiseuille profile adds a quadratic, pressure driven, component to the Couette contribution, here with non-vanishing mean~advection. \label{f2}}
\end{figure}

In the first scenario, at increasing $\Rey$, the base state is continuously changed into more and more complex flow regimes resulting from a cascade of instabilities ending in turbulence (Figure~\ref{f1}a).
Importantly, this scenario develops from a linear primary instability amplifying infinitesimal disturbances beyond some threshold $\Rec$.
The subsequent cascade involves a finite number of steps, while at each step, the bifurcated and bifurcating states exchange themselves as $\Rey$ is varied.
The cascade is essentially reversible with no (or very limited) hysteresis as $\Rey$ is swept up and down, a property best conveyed by the expression \emph{globally super-critical}.~A typical closed flow example is convection in a horizontal fluid layer originally at rest and heated from below with differential buoyancy playing the destabilizing role, see \S3.2.2 in \cite{Ma10}.
This scenario is relevant every time the dynamics away from the base state can be analyzed using the standard tools of linear stability analysis and weakly nonlinear perturbation theory, at least in principle since technical difficulties can be insurmountable beyond the few first steps.
 This is of course the case for convection, but also for open flows with velocity profiles displaying inflection points (unstable according to Rayleigh's inviscid criterion~\cite{RaXX1}; see~\cite{Ma10,SH01}), e.g., a shear layer downstream a splitter plate (Figure~\ref{f2}a) or a wake downstream a blunt obstacle, in which case primary destabilization arises from a Kelvin--Helmholtz instability, while viscosity plays its intuitive stabilizing role on a primary instability that develops at low Reynolds~number (\S7.2.2 in \cite{Ma10}).

In this review, I will be concerned with the alternative \emph{catastrophic} scenario in which nonlinearity plays the essential role, while dissipative processes are less efficient in controlling the flow.
 (To my knowledge, the term `catastrophic' was introduced by Coles~\cite{Co6x2} who described the first scenario as `spectral evolution', an expression that conveys the right idea, but is less nonlinearly connoted than `globally super-critical'.) 
Linearity is associated to uniqueness of solutions, namely the laminar base flow response to small driving away from equilibrium along the \emph{thermodynamic} solution branch.
On the other hand, far from equilibrium, nonlinearity indeed permits a multiplicity of solutions to the NSE.
The transition to turbulence is now much wilder, laminar flow directly competing with a turbulent regime and no stage of intermediate complexity in between (Figure~\ref{f1}b).
Coexistence of locally stable solutions being the mark of sub-criticality in elementary bifurcation theory, this scenario can legitimately be termed \emph{globally sub-critical} \cite{MK05,Gr00,Ma16a}.
It also displays strong hysteresis upon sweeping $\Rey$ up and down and, 
on general grounds, a \emph{global stability threshold} $\Reg$ can be defined, corresponding to the value of $\Rey$ below which the base state is unconditionally stable.
Sustained coexistence can accordingly be observed for~$\Rey\ge\Reg$.

The \emph{phase space} interpretation of multi-stability is straightforward in \emph{confined} systems   where lateral boundary effects enforce the spatial coherence of nonlinear modes (see \S3.3.2--4 in \cite{Ma10}).
Confinement effects are appreciated through aspect ratios, viz. $\Gamma=L/\lambda_{\rm c}$ where $L$ is a typical extrinsic scale of interest and $\lambda_{\rm c}$ the intrinsic scale generated by the instability mechanism, i.e., $\approx$ the number of cells in convection.
Open flows through pipes or along plates are always extended at least in the streamwise direction.
By placing artificial periodic boundary conditions at small distances, the ensuing \emph{low-dimensional} dynamical-system reduction undoubtedly helps one identifying locally relevant nontrivial solutions in \emph{phase space}~\cite{Getal08,Ketal12}, but does not provide any understanding of the coexistence of such local solutions with the laminar flow solution in different regions of \emph{physical space}, which, as earlier stressed by Pomeau~\cite{Po86}, is the prominent feature to explain:
in the catastrophic case, chaos is \emph{spatiotemporal} in essence, and the whole system of interest is better viewed as a patchwork of subdomains filled with either laminar or turbulent flow separated by sharply-defined interfaces.
Due to intrinsic stochasticity in the local nontrivial state, these interfaces permanently fluctuate, generically leading to a regime of spatiotemporal intermittency~\cite{CM94}.~Though still conceptually appealing, the phase-space picture, mostly valuable for confined systems, becomes unpractical and, possibly, even misleading.
Natural observables are now statistically-defined quantities such as the \emph{turbulent fraction}, the mean fraction of space occupied by the turbulent state (spatial viewpoint) or the \emph{intermittency factor}, the mean fraction of time spend in the turbulent state (temporal viewpoint) and higher spatiotemporal statistical moments defined via the laminar/turbulent~dichotomy.

I now turn to examples taken from planar configurations where a fluid flows along solid boundaries for which, controlled by viscosity, the velocity profile is deprived from inflection point, Rayleigh's inviscid condition for linear stability~\cite{RaXX1}, and remains stable up to large Reynolds numbers (Figure~\ref{f2}b--d).
This is the case of the \emph{channel} flow between parallel plates under pressure gradient with the parabolic$\,$ 
Poiseuille profile (PPF), of the \emph{simple shear} flow between counter-sliding parallel plates with linear Couette profile (PCF; Figure~\ref{f2}c), of a mixture of the two with a more general quadratic profile called Couette--Poiseuille flow (CPF; Figure~\ref{f2}d), 
of the plane \emph{boundary layer} flow in the absence of pressure gradient with Blasius profile (BBL; Figure~\ref{f2}b), or its variant with permeation at the wall, the asymptotic suction boundary layer (ASBL).
Flow under pressure gradient in \emph{pipes} of circular section with parabolic Poiseuille profile, the Hagen-Poiseuille flow (HPF), or nearly square section, also enter this category \cite{Betal15}.

In all of these cases, viscous effects do not play their simple, low-$\Rey$, damping role, but a more subtle part in a mechanism producing Tollmien--Schlichting (TS) waves, effective at high $\Rey$ only~\cite{SH01}.
Among the cases mentioned above, PPF, BBL and ASBL, have finite, but high enough TS-threshold $\Rec$, while PCF is known to be linearly stable for all~$\Rey$ and HPF believed to be so.
These two, PCF and HPF, therefore come out as paradigms of systems controlled by mechanisms that do not rely on conventional linear stability analysis, such as the Kim--Hamilton--Waleffe self-sustainment process (SSP) \cite{HKW95,Wa97,MK05}.
The SSP is a cyclic process where perturbations in the form of low-level \emph{streamwise vortices} induce by \emph{lift-up} large spanwise modulations of the base flow called streamwise \emph{streaks}. When sufficiently amplified, these streaks are themselves unstable via the development of locally inflectional velocity profiles, provoking their breakdown. In a third step breakdown products are filtered out to regenerate the streamwise vortices \cite{HKW95,Wa97}. For an illustration, see Gibson's video \emph{Turbulent dynamics in a `minimal flow unit'} on {\tt channelflow.org} \cite{ChFl}, choosing tab `{\tt movies}' among the headings.
Nontrivial solutions brought about by such inherently nonlinear couplings can then be found away from the base flow in an intermediate $\Rey$ range, $1\ll \Rey \ll \Rec$ (possibly infinite).
Scenarios resting on the presence of a linear instability (infinitesimal disturbances) are, in practice, \emph{bypassed} by the amplification of \emph{finite}-amplitude, \emph{localized} perturbations pushing the flow in the attraction basin of these nontrivial states living on the turbulent solution branch.
As previously mentioned, this branch will be stable for $\Rey\ge\Reg$, but its states are only transient below.
Now, on general grounds, a regime of spatially uniform or \emph{featureless} turbulence~\cite{Aetal86}, is expected at very large $\Rey$ with turbulent fraction or intermittency factor saturating at one~\cite{Co6x1,Co6x2}.
On the other hand, just above $\Reg$, one may expect these quantities to be markedly smaller than one, characterizing the conspicuous laminar-turbulent alternation.
How do they approach saturation as $\Rey$ increases, either through a smooth crossover or at a well-defined upper threshold $\Ret$, and more generally, how do they vary all along the \emph{transitional range} between $\Reg$ and the putative $\Ret$ are the questions of interest.

Before discussing the \emph{two-dimensional} (2D) transitional regime for flows along plates with laminar-turbulent patterns depending on two directions, streamwise and spanwise, I now briefly review the other paradigmatic case considered first by Reynolds in a transition perspective~\cite{Re83}, namely HPF, the flow along a straight pipe (hence \emph{one-dimensional}, 1D for short).
This summary is just given for further reference since several extensive accounts can be found in the recent literature~\cite{Eetal07,Mu11} and one can rely on the remarkable article by Barkley~\cite{Baxx2} for a brilliant analysis of theoretical issues and associated modeling.
First, the parabolic HPF profile is presumed to be linearly stable for all $\Rey$, whereas at moderate $\Rey$, once triggered, turbulence remains localized in isolated coherent chaotic \emph{puffs} with finite lifetimes that increase super-exponentially with $\Rey$~\cite{Hetal06}.
Transient turbulence happens to become sustained because, when $\Rey$ gets larger, before decaying puffs can split and propagate localized chaotic disturbances further, thus contaminating the flow.
The threshold $\Reg$ can be defined without ambiguity when decay is statistically compensated by splitting so that turbulence persists on average~\cite{Aetal11}.
Somewhat above $\Reg$, turbulence inside the puffs becomes more aggressive and the puffs turn into turbulent plugs called \emph{slugs} that, when $\Rey$ increases a bit, grow in the upstream direction despite downstream advection~\cite{Betal15}. As $\Rey$ further increases, laminar-turbulent intermittency is progressively reduced to the benefit of featureless turbulence, hence a smooth crossover rather than a threshold at some well defined $\Ret$.~At the phenomenological level, a remarkably successful model covering the whole transitional regime has been developed by Barkley~\cite{Baxx1,Betal15,Baxx2}.
The reaction-diffusion-advection process \cite{Mu93} in terms of which this model is formulated will have some relevance to the discussion of the 2D organization of the laminar-turbulent coexistence~(Section \ref{S6}).

At this stage, as a step toward the problem of 2D patterns proper, I should point out that the transition from 1D axial to 2D wall-parallel dependence of the turbulence intensity can be studied in a few experimental settings with straightforward numerical implementation.
Annular Poiseuille flow, the flow between two coaxial cylinders driven by a pressure gradient, is a first example.
When the radius ratio is small, despite the presence of the inner cylinder, the transition to/from turbulence basically follows the 1D scenario of pure HPF with no inner cylinder (puffs, slugs, etc.).
On the other hand in the small gap limit, when this ratio tends to one and the curvature of the fluid layer tends to zero, a laminar-turbulent organization takes place in the form of turbulent helices translating in the gap, locally anticipating the oblique bands of the planar case.
The crossover between the two regimes has been studied as a function of the radius ratio \cite{Ietal16,IT17,Ietalx}.
In the same way, fluid motion induced by steadily counter-sliding the cylinders along their axis, yields annular Couette flow (more easily implemented numerically~\cite{Ketal16c} than approximated experimentally) that connects to PCF in the small gap limit in the same way as annular to plane Poiseuille flow.
In the next section, I turn to the Couette case but when the two cylinders are differentially rotating around, rather than translating along, their common axis in the moderate-to-small gap range for which the local state, once bifurcated, depends on axial and azimuthal coordinates from the start, being thus genuinely~2D.

\section{Cylindrical Couette Flow\label{S2}}

Together with the flow through a pipe and thermal convection, cylindrical Couette flow (CCF), represents one of the most emblematic testbeds for studying hydrodynamic stability and the transition to turbulence~\cite{Fetal14}.
This rich experimental configuration is geometrically specified by the radii $r_{1,2}$ of the cylinders (inner$\,$: 1, outer$\,$: 2), ratio $\eta=r_1/r_2$ measuring curvature effects, the axial and circumferential aspect ratios, $\Gamma_z=L/d$ and $\Gamma_\theta=\pi (r_1+r_2)/d$,\quad $d=r_2-r_1$ being the gap between the cylinders and $L$ their length (usually fairly large when compared to $d$), and the rotation rates $\Omega_{1,2}$.
By convention, $\Omega_1\ge0$ with $\Omega_2=0$, $>0$, or $<0$ for the outer cylinder at rest, co-rotating, or contra-rotating with respect to the inner cylinder, respectively.
One usually defines the inner and outer Reynolds numbers as $\Rey_{1,2} = \Omega_{1,2} r_{1,2} d/\nu$, ($\nu\,:$ kinematic viscosity), but other physics-motivated parameterizations are possible, such as the Taylor number~\cite{Ta23}. A definition referring to the mean shear,
 $\overline\Rey=(\Rey_1-\eta \Rey_2)/2(1+\eta)$, is particularly helpful for direct comparisons with other wall-bounded configurations~\cite{BTxx1,BTxx2,Ma04,TB11}, especially plane Couette flow in the limit $\eta\to1$.
The advantage of CCF is that most situations of interest can be spanned~\cite{Fetal14}, from temporal chaos (short cylinders, wide gap) to spatiotemporal chaos (gap small compared to perimeter), and from globally super-critical to globally sub-critical according to whether or not the dynamics is controlled by the centrifugal instability of the innermost fluid layer at the inner cylinder \cite{RaXX2,Ta23}.
At this point, I want to stress that the present review is restricted to the globally sub-critical transitional regime where laminar-turbulent patterns form.
I will not consider the fully developed regime much beyond the limit for featureless turbulence~\cite{Getal16} and, apart from a brief mention below, I will not consider the globally super-critical case in detail, leaving it to~\cite{Aetal86,DS85,Fetal14}.

CCF is entrained by the motion of the cylinders where no-slip conditions apply.
All along the thermodynamic branch, the \emph{base state} displays a purely azimuthal velocity profile, entirely controlled by viscous effects.
In the inviscid case, $\nu=0$, when the Rayleigh stability criterion is violated --here, when the angular momentum does not increase monotonically outwards~\cite{RaXX2}-- infinitesimal perturbations to the base flow are amplified through inertial effects while a finite viscosity delays the instability until a shearing threshold is reached.
A super-critical instability then develops producing axisymmetric Taylor vortices~\cite{Ta23}. 
This is the case when the Rayleigh criterion for stability is violated all over the gap, i.e., $0\le\Omega_2\le\eta\Omega_1$.
Taylor instability is then at the start of a globally super-critical sequence of bifurcations toward more and more complicated flow behavior up to a turbulent regime, a scenario termed `spectral evolution' by Coles~\cite{Co6x2} who early reported on it.
Consult~\cite{DS85,Fetal14} for reviews and Figure~1 in \cite{Aetal86} for a detailed bifurcation diagram at $\eta=0.883$.
The typical wavelength of Taylor rolls is twice the gap and when the axial aspect-ratio $\Gamma_z$ is small enough, the setup accommodates a small number of rolls that remain highly coherent even when the flow enters the turbulent regime, then rather understood in terms of  \emph{temporal} chaos within the theory of low-dimensional dissipative dynamical systems (see Chapter~4 in \cite{Ma10}).
When $\Gamma_z$ is large, CCF can be studied using the  \emph{envelope} and  \emph{phase} formalisms, turbulence acquiring a more  \emph{spatiotemporal} flavor~\cite{Br88}, still in a globally super-critical context.

When the two cylinders rotate in opposite directions, $\Omega_2<0$, the Rayleigh criterion for stability is violated only in a fluid layer near the inner cylinder where unstable linear modes with non-axisymmetric structure can develop~\cite{Aetal86}.
Near the outer cylinder, the criterion is fulfilled so that the corresponding fluid layer is stable in the inviscid limit, right in the situation described above for globally sub-critical plane flows.
Localized finite amplitude perturbations bursting from the inner unstable layer~\cite{CM96} can now trigger the transition to turbulence.
Bursting perturbations affect a network of interpenetrating spirals (IPS) \cite{Aetal86} generating \emph{turbulent spots}, at first intermittent and disseminated, but more and more persistent as the shear increases~\cite{Co6x1,Co6x2}.
Turbulent spots further grow into turbulent patches and next into \emph{spiral turbulence} (ST regime), characterized by its helical, \emph{barber pole}, aspect first reported by Coles \cite{Co6x1}, later scrutinized by Andereck et al.~\cite{Aetal86} and others, e.g. \cite{Hetal89,LR98}.
Upon further increasing the shear, the helical arrangement disappears above some mean shear threshold $\overline\Rey_{\rm t}$, translated as a line in the $(\Rey_2,\Rey_1)$ parameter plane, beyond which the flow enters the featureless turbulent (FT) regime, thus saturating the turbulent fraction or equivalently the intermittency factor (line $\gamma=1$ in Figure 2a,b of \cite{Co6x2}; see also Figure~1 of \cite{Aetal86}).
A direct collapse of turbulence to axisymmetric laminar flow can be observed for very fast counter-rotation as a direct transition in the `hysteresis region' in Figure~2a of \cite{Co6x1} or Figure~3 of \cite{VA66}, with features specific to transient temporal chaos when $\Gamma_z$ is small, enforcing spatial coherence~\cite{Betal10}.
At more moderate counter-rotation rate, decay happens {via} IPS in the shear range just before axisymmetric laminar flow is recovered (Figure~1 of \cite{Aetal86}).

In the experiments mentioned above, all with $\eta\sim0.88$ ($\Gamma_\theta=50$) a single helical branch (\mbox{$n_\theta=1$}) was ordinarily obtained \cite{Aetal86,Co6x1,Co6x2}.
Thinking in terms of a laminar-turbulent pattern, owing to azimuthal periodicity, a single helix branch corresponds to an oblique band and, accordingly, a streamwise wavelength $\lambda_\theta=\Gamma_\theta d/(n_\theta\!\!=\!\!1)=50d$; 
not currently observing $(n_\theta=2)$-helices means $\lambda_\theta>\Gamma_\theta d/2\approx 25d$, which is confirmed by the fact that no pattern was found for $\eta\le0.75$, i.e., $\Gamma_\theta\le 22$. 
Patterns with wavelengths very large when compared to the gap $d$ are therefore observed. 
In order to approach the paradigmatic case of PCF, experimental configurations with $\eta$ closer to 1 have been considered.
Prigent~\cite{Petal03,PD05} scrutinized the cases $\eta=0.963$ and $0.983$, hence $\Gamma_\theta=167$ and $358$.
Besides noting a continuous shift of the bifurcation diagram towards the $(\Omega_2 = -\eta\Omega_1)$-line in the $(R_2,R_1)$-parameter plane as $\eta$ approached the PCF limit $\eta=1$, he obtained helices with more branches and wavelengths $\lambda_\theta=\Gamma_\theta d/n_\theta$ in agreement with those for $\Gamma_\theta\sim50$ and $n_\theta=1$.
A few supplementary features are worth mentioning.
(i) In all cases, the spiral patterns appeared to be nearly at rest in a framework rotating at the mean angular speed $(\Omega_1+\Omega_2)/2$~\cite{Co6x1,Co6x2,Petal03}; 
({ii}.a) The helical pattern emerged continuously from the FT regime with, close to $\Ret$, domains of opposite-helicity modes separated by grain boundaries (Fig.9 in \cite{Petal03}) seen to move so as to favor a single helicity farther from $\Ret$.
({ii}.b)~In the single-helicity regime, the azimuthal and axial wavelengths were seen to vary with the mean shear, with larger wavelengths close to decay at $\Reg$ (Figure~5 of \cite{Petal03});
(iii) When the pattern was well established, the laminar-turbulent interfaces displayed \emph{overhangs}, that is, quiescent flow close to one cylinder facing turbulent flow near the other~\cite{Co6x1,Co6x2,VA66,Do09}.
I will come back to the emergence of the spirals at $\Ret$ from the featureless regime and their characterization in~Section \ref{S5}.

\section{The Laminar-to-Turbulent and Turbulent-To-Laminar Transition in Planar Flows\label{S3}}

In this section, I first present the general features of laboratory and numerical studies for PCF achieved by shearing a fluid between two parallel plates moving in opposite directions, conceivably the simplest possible planar shear flow, before considering other standard flow configurations such as plane channel flow, in their relation to laminar-turbulent patterning.
I defer the general question of turbulence breakdown at $\Reg$ to Section \ref{S4} and pattern emergence at $\Ret$ to Section~\ref{S5}.

\subsection{Plane Couette Flow\label{S3.1}}

Ideal PCF can be entirely characterized by the Reynolds number $\Rey^{\rm PCF}=V h/\nu$ where the plate speed $V$ serves as speed scale.
Usual conventions for PCF are that Plate 2 at $y=y_2$ slides in direction $x$ (streamwise), with speed $V_2=V>0$, and Plate 1 at $y_1<y_2$ with speed $V_1=-V<0$.
The half-gap $h=(y_2-y_1)/2$ is usually taken as length unit and $\nu$ is again the kinematic viscosity (Figure~\ref{f2}c).
$\Rey^{\rm PCF}$ is nothing the Reynolds number  $\overline\Rey$ based on the mean shear for CCF defined earlier with $|\Omega_2|=\eta\Omega_1$ in the limit  $\eta\to1$.
Concrete experimental or numerical realizations require specifications of the system size via aspect ratios, streamwise $\Gamma_x=L_x /2h$ and spanwise $\Gamma_z=L_z/2h$.

From an experimental point of view, most often the fluid is driven by an endless belt forming a closed loop entrained by two cylinders~\cite{Re59}.
In the experimental configuration now generally considered, good control of the gap is obtained by the addition of guiding rollers~\cite{TA92,Detal92}.
Early experiments ca.~1960 were mostly dedicated to the statistical properties of the fully turbulent regime~\cite{Re59}.
Transitional issues only began to be considered at the beginning of the 1990's with works in Stockholm (Sweden)~\cite{TA92,Betal95} and Saclay (France)~\cite{Detal92,DD95}.
Like in other planar flows, laminar-turbulent coexistence first manifests itself in the form of \emph{turbulent spots}.
Since laminar flow is linearly stable at the considered $\Rey$, they must be \emph{triggered} by localized finite-amplitude perturbation of the laminar flow.
The shape and strength of the perturbations necessary to obtain growing spots were studied as functions of $\Rey$, with the result that the higher $\Rey$, the smaller the perturbations need to be, while for large enough $\Rey$ low level background turbulence is sufficient to promote the transition.

The global stability threshold $\Reg$ was first qualitatively located around $\Rey=360$ using a simple growth-or-decay criterion for spots~\cite{TA92}.
More quantitative results were obtained from the divergence of the spots' mean lifetimes for $\Rey<\Reg$~\cite{Detal92}, or of the perturbation's amplitude necessary to promote the transition for $\Rey>\Reg$~\cite{DD95}.
It was soon recognized that the problem was statistical in essence, since at given $\Rey$ above $\Reg$, not every triggering was successful, but only a fraction, the larger the higher above $\Reg$.
Like for CCF, the first experiments were performed with relatively large gaps, hence small aspect ratios, typically $L_x\times L_z=300h\times80h$~\cite{TA92,Detal92,Betal98}.
Longer observation times and better statistics in larger domains ($570h\times140h$) led to decrease the estimate down to about $325$, confirmed by experiments where a turbulent regime at high $\Rey$ is suddenly \emph{quenched} at a final value around $\Reg$~\cite{Betal98}.
At threshold, the steady-state turbulent fraction was seen to drop discontinuously to zero, while long transients with well-defined turbulent fraction were observed before decaying to laminar flow in the long term~\cite{Betal98}.
Growing spots were followed over longer durations, reaching a mature stage (Figure~1 (right) of \cite{Betal98}), with random splittings and recombinations, and showing a trend to stationary, oblique patches of both orientations w.r. to the streamwise direction (Figure~2 of \cite{BC98} or Figure~3 of \cite{MD01}).
For reviews of the Saclay results, consult \cite{MD01,PD05}.

Paralleling his study of CCF, Prigent systematically focused on a regular oblique banded regime obtained in a larger PCF setup ($770h\times340h$) by slowly decreasing $\Rey$ from high values where the flow is uniformly turbulent (featureless) \cite{Petal03,PD05}.
The pattern was observed below $\Ret\approx415$, but neatly organized only below $\Rey=402$.
The two possible orientations were present in the form of chevrons at $\Rey=393$ and a single one from $\Rey\approx380$ down to about $350$.
Below, the whole pattern was broken into large domains of opposite orientations separated by grain boundaries resembling what was observed earlier at smaller aspect ratios \cite{MD01}.
This sequence is illustrated in panels a--c of Figure 3 in \cite{Petal03}.
Below $\Rey=325 \sim \Reg$ the flow was again found laminar.
The pattern's wavelength was observed to stay roughly constant in the streamwise direction ($\lambda_x\approx100h$), but the spanwise wavelength $\lambda_z$ was seen to vary from $50h$ close to the featureless regime up to $85h$ close to breakdown.
Recalling that $d=2h$, comparable wavelengths were obtained for CCF at $\eta=0.983$~(Figure~5 in \cite{Petal03}) and, once expressed in terms of mean shear Reynolds number $\overline\Rey$ introduced earlier, thresholds $\Ret$ and $\Reg$ were in close correspondence \cite{Ma04}.

Numerical simulations have considerably contributed to our empirical knowledge of transitional wall-bounded flows in general and PCF in particular.
They have been systematically developed in parallel with laboratory experiments from the turn of the 1990s, starting with Lundbladh and Johansson's~\cite{LJ91} on spot growth in an extended periodic domain.~At the same time, Jim\'enez and Moin~\cite{JM91} introduced the concept of \emph{minimal flow unit} (MFU), a periodic domain virtually confined by periodic boundary conditions placed at streamwise ($\ell_x$) and spanwise ($\ell_z$) distances small enough that low-$\Rey$ driving is just able to maintain nontrivial flow.
This concept was first adopted to educe the SSP by Hamilton~et~al.~\cite{HKW95,Wa97} who found values as small as $\ell_x\simeq5.5h$ and $\ell_z\simeq3.8h$ for PCF.
The same setting next served to identify a number of exact solutions to the NSE in the spirit of low-dimensional dynamical systems theory~\cite{Eetal08,Ketal12}.

Considering large aspect-ratios of interest to the study of laminar patterning, early work related either to the evolution and growth of turbulent spots~\cite{LJ91} or to the developed stage at moderate $\Rey$, but largely above $\Ret$~\cite{Ketal96}.
Fully resolved computations dedicated to the transitional range are more recent, owing to the numerical power needed.
For example Duguet et al.~\cite{Detal10} obtained results in general agreement with laboratory experiments (thresholds, wavelengths).

To circumvent the high computational cost of well-resolved simulations, Barkley and Tuckerman chose to treat the actual three-dimensional problem as solved in a narrow and elongated (quasi-1D) oblique domain with an orientation fixed in advance and helical boundary conditions cleverly-chosen to mimic the in-plane 2D part of the actual 3D problem by periodic continuation (Figures 1 and 2 in \cite{BTxx1}).
The helical condition was sufficient to deal with streamwise correlations essential to reproduce the main characteristics of the bands~\cite{PM11}.
Among properties of the laminar-turbulent patterning, Barkley and Tuckerman~\cite{BTxx1,BTxx2,TB11} analyzed the structure of the mean flow inside the laminar bands, something hardly detectable in early laboratory experiments, but considered later~\cite{CM15}.
Despite forbidding an account of orientation fluctuations near $\Ret$ (see Section \ref{S5}), this approach had deep influence on subsequent research~\cite{Setal13,Tetal09,Tetal14}.
\begin{figure}

\includegraphics[angle=90,width=0.24\textwidth]{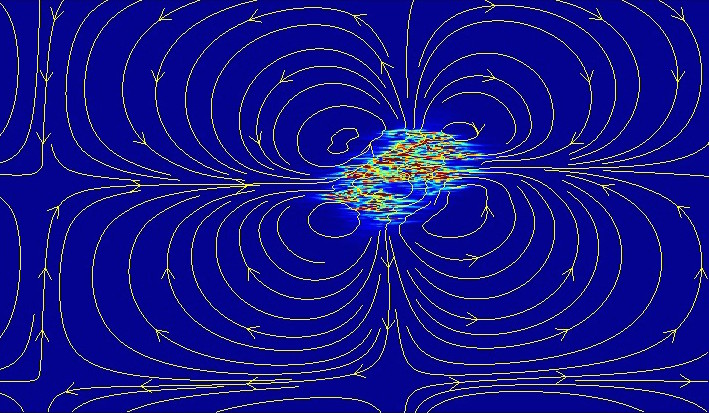}\hfill
\includegraphics[angle=90,width=0.24\textwidth]{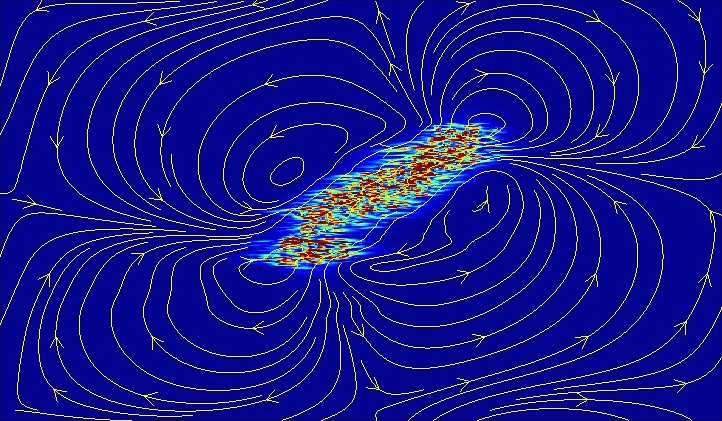}\hfill\includegraphics[angle=90,width=0.24\textwidth]{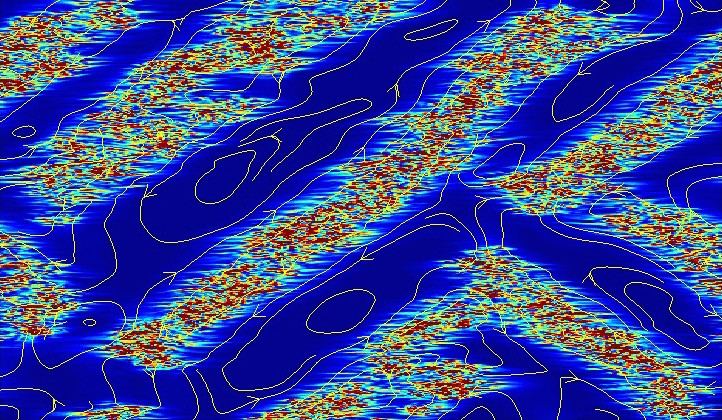}\hfill
\includegraphics[angle=90,width=0.24\textwidth]{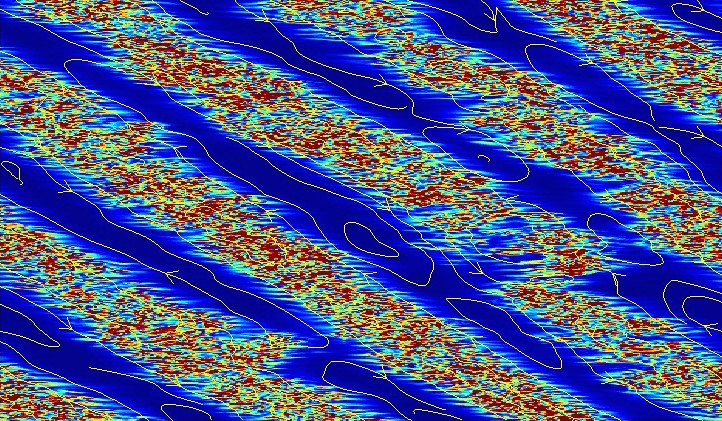}

\caption{Laminar-turbulent patterning in PCF modeled via under-resolved direct numerical simulations in a $468h\times272h$ domain (see \cite{MFDR} for details).
Snapshots of typical flow states using color levels of the perturbation energy averaged over the gap $2h$: deep blue is laminar, but different from the base flow due to the large-scale mean flow component featured by the faint yellow lines on the blue background.
The streamwise direction is vertical.
From left to right: Turbulent spot at $\Rey=281.25$, will decay after a very long transient. Growing oblique turbulent patch for $\Rey=282.50$. Mature, but unsteady pattern with statistically constant turbulent fraction at $\Rey=283.75$. Well-formed pattern with larger turbulent fraction at $\Rey=287.50$. Values of $\Rey$ are shifted downward due to modeling via under-resolution \cite{MR11}, but the pictures displayed give a good idea of experimental findings described at the beginning of the~section. \label{f3}}
\end{figure}

Another way to look at large aspect-ratios while limiting the numerical demand is by degrading the numerical resolution (Figure~\ref{f3}).
The idea is that, at the intermediate Reynolds numbers of interest in transitional studies, it is sufficient to render the coherent structures at the scale of the gap between the plate while accepting that the smallest wall-normal scales close to the solid plates be only approximately evaluated.
 This can be turned into a systematic modeling strategy~\cite{MR11} and, in practice, 
bands appear to be an extremely robust feature of the transitional regime since reliable hints about the local processes involved in the growth and decay of the pattern around $\Reg$ can be obtained in this way~\cite{MFDR}, the price to be paid being a systematic decrease of $\Reg$ and $\Ret$ (also observed in other cases such as ASBL \cite{Ketal16a}), which can be explained by a default of dissipation in the smallest scales rendering the flow more turbulent than it should be at given $\Rey$.

The robustness of band patterning is also illustrated by more drastic modeling options, in particular by changing the boundary conditions from no-slip to stress-free, hence simplifying the analysis while trying to keep the physics of the problem, much like in Rayleigh's early analysis of convection~\cite{RaXX3}.
Soon after Waleffe's modeling effort illustrating the SSP within the MFU framework \cite{Wa97}, I extended the approach to the spatiotemporal domain~\cite{MD01}.
Strong support to this practice has recently been given by Chantry et al. \cite{Cetalxx1}, who further put forward the idea that, at the price of an empirical length rescaling, one could map the no-slip and stress-free problems onto each other.
An `interior flow' could then be defined by a matching of flow profiles at a statistical level apart from layers close to the plates, a procedure that could next be applied to other globally sub-critical flows of interest~\cite{Cetalxx1}.
The crucial point is next that reducing the wall-normal expansion to very few modes is sufficient to account for the most relevant characteristics of the flow, in particular the patterning~\cite{Ma16b}.
Stress-free boundary conditions allow the use of trigonometric functions~\cite{RaXX3} that greatly ease the exact analytical treatment and the subsequent work-load reduction \cite{MD01,Cetalxx1}, but this reduced description is not limited to the stress-free case: comparable results can also obtained in the no-slip case with adapted basis functions~\cite{LM071,LM072,SM15}, or for possibly other systems in the same class.
The no-slip approach is more cumbersome, but has the merit to make the structure of the resulting model explicit, and to point out that specificities of the problem only lie in the precise values of the coefficients in the general model~\cite{SM15}.
On another hand, the considerable reduction implied in the stress-free modeling helped \mbox{Chantry et al. \cite{Cetalxx2}} to consider the decay of turbulence in very large systems as discussed later in~Section \ref{S4}.
I come back to modeling issues in Section \ref{S6}.

\subsection{Other Planar Configurations\label{S3.2}}

PCF considered above was achieved between two walls moving at the same speed in opposite directions~\cite{TA92,Detal92}, hence no mean advection and coherent structures nearly at rest in the laboratory frame.~Similar results are obtained in configurations where additional effects are introduced, for example Coriolis forces when the setup is placed on a rotating table~\cite{Tetal10,Betal12}.~Whereas anti-cyclonic rotation, i.e., opposed to the rotation induced by the shear, is destabilizing and yields a globally super-critical situation comparable to that of co-rotating Taylor--Couette flow, cyclonic rotation is stabilizing and exacerbates the globally sub-critical character of the transition so that a laminar-turbulent sequence similar to that in non-rotating PCF is obtained, thresholds increasing roughly linearly with the rotation rate \cite{Betal12,Tetal10}.
In the upper transitional range, above the oblique band regime and close to $\Ret$, an interesting regime called `intermittent' is observed, closely resembling what is observed in the corresponding range of CCF~\cite{Petal03}.

Another way to achieve PCF is by moving a single wall, with the other one at rest.
In an open configuration~\cite{Re59}, perturbations to the base flow would be advected downstream, but in a closed configuration \cite{Ketal17}, the fluid entrained by the wall tends to accumulate at the downstream dead-end and an adverse pressure gradient builds up, adding a parabolic Poiseuille component to the initial linear Couette profile thus achieving a Couette--Poiseuille flow (CPF) profile (Figure~\ref{f2}d), here with zero mean advection at steady state.
This CPF profile is also linearly stable for all $\Rey$ and prone to a direct transition to turbulence as recently shown by Klotz et al. \cite{Ketal17} who observed turbulent spots evolving into a~steady oblique turbulent band as $\Rey$ increased.
Comparable to that of early PCF experiments~\cite{TA92,Betal98}, the experimental aspect ratio was too small to allow the observation of several bands.

As far as the transition to/from turbulence is concerned, the study of CPF is recent when compared to that of PPF, a flow configuration for which the dynamics of turbulent spots has been early studied in detail~\cite{CWP82}.
See~\cite{Letal14} for a recent investigation dedicated to mechanisms for spot development in relation with Barkley's puff/slug sustainment mechanism~\cite{Baxx1,Baxx2}.
In close parallel with the case of PCF, pattern formation along the transitional range at large aspect ratios has been examined only recently.
The earliest report of oblique patterning is from Tsukahara's group, both in numerical simulations~\cite{Tetal05} and in laboratory experiments~\cite{Hetal09}.
This observation was completed by several other groups who obtained sustained oblique short band fragments or isolated bands at very low~$\Rey$, numerically \cite{TI14,Xetal15,Ketalxx,Petalxx} as well as experimentally \cite{Petalxx}.
These localized solutions are present at values of $\Rey$ clearly lower than for sustained or growing standard turbulent spots \cite{CWP82,Letal14} or for pattern decay in large aspect-ratios systems using the experimental protocol of Sano and Tamai \cite{ST16}.
Their role in the transition process thus remains to be elucidated.
Interestingly, the obliquely-patterned transitional range could also be reproduced by \emph{a priori} fixing the orientation in a Barkley--Tuckerman elongated computational domain~\cite{Tetal14}.
The approach showed in particular that the bands slowly move with respect to the mean flow, slower at higher $\Rey$ and faster at lower $\Rey$, in contrast with PCF where bands are essentially at rest in the laboratory frame for symmetry reasons. 

As mentioned earlier, for annular Poiseuille flow in the limit of radius ratio tending to one, the pattern is, not unexpectedly, in the form of intertwined helices~\cite{Ietal16,Ietalx,IT17}.
In the planar case, the effects of spanwise-rotation on thresholds and pattern wavelengths have also been studied~\cite{Ietal15}, but their variations turn out to be more complicated than for PCF due to a different shear wall-normal~dependence.

The list above is not limitative and laminar-turbulent patterning can be observed in numerous other systems, such as plane Couette flow in the presence of a density stratification imposed by a stabilizing temperature gradient~\cite{Detal15}.~The Ekman boundary layer close to a rotating wall in a stably-stratified fluid is an other example with banded turbulence present when the density stratification is sufficiently strong \cite{Detal14}.~In the case of torsional Couette flow, the flow between two plates rotating around a common perpendicular axis \cite{Letal10}, with a shearing rate depending on the distance to the rotation axis and the differential rotation rate, several regimes can be observed in a single experiment, from scattered spots to a laminar-turbulent arrangement of spiral arms \cite{CLG02}.
Other systems, possibly more easily defined in a numerical-simulation context than really achievable in a laboratory environment, are also of interest with respect to the role of additional stabilizing forces on patterning \cite{Betal12}.

Generalizing the remarks at the end of Section \ref{S2} one should notice that, once the physically relevant velocity and length scales are identified and the appropriate Reynolds number is defined (\S7.3.4 in \cite{Ma10}), intervals $[\Reg,\Ret]$ for all of these systems fall in the same range at least in order-of-magnitude~\cite{Ma04,BTxx1,BTxx2,TB11}, and even quite close as in the case of PCF and CCF with $\eta$ tending to one.
A second other common feature is that the patterns' wavelengths are very large when compared to this most relevant length scale, which immediately raises questions, still mostly open, about the mechanisms controlling the spatial periods of the laminar-turbulent alternation and its orientation with respect to the streamwise direction.
Third, they travel in the system at a speed that is very close to that of the mean flow rate, i.e., at rest in the laboratory frame for PCF or CPF without mean advection.

The standard Blasius boundary layer along a flat plate (Figure~\ref{f2}b), so important in applications, has not been, and will not be, considered here, despite the fact that the expressions `intermittency factor' or `turbulent fraction' have been coined to deal with its transitional behavior; for an early review with illustrations, see \S{}D of \cite{Co6x1}.
This is because the natural transition strongly depends on the quality of the in-flow and its globally sub-critical character is less marked.
In terms of distance to the leading edge, the linear Tollmien--Schlichting modes indeed become unstable soon after the nonlinear catastrophic turbulent-spot bypass that rapidly leads to a fully turbulent boundary layer without any intermediate pattern stage.
See~\cite{Ketal16b} for an interesting recent approach to this transition with references to earlier work.

The developing character of the boundary layer can however be suppressed by applying a uniform suction, i.e., a through-flow at the plate taken as porous, as achieved experimentally by Antonia~et~al.~\cite{Aetal88}.
The so-called asymptotic suction boundary layer (ASBL) then becomes independent of the downstream distance and can be characterized by a Reynolds number $\Rey$ that is just the ratio of the velocity of the fluid far away from the plate $U_{\infty}$ to the suction velocity $V_{\rm s}$.
Its simplicity make it convenient to numerical study.
The main result of Khapko et al. \cite{Ketal16a} pertaining to the turbulent pattern formation problem is that at $\Reg$ the boundary layer abruptly decays without showing oblique laminar-turbulent interfaces.
This observation was linked to the unbounded character of turbulent perturbations in the wall normal-direction, thought to impede the formation of any wall-parallel large-scale secondary flow able to maintain the obliqueness of interfaces~\cite{DS13}.
This argument was further supported by the restoration of such an obliqueness when a virtual wall was placed parallel to the plate by artificially damping perturbations beyond some distance to it.
The same argument holds for the stratified vs. unstratified Ekman boundary layer, when a sufficiently strong stratification naturally exerts the appropriate damping.
This remark on a possible key ingredient for patterning closes the empirical part of this review, the rest of which is devoted to studies more related to the qualitative and quantitative understanding of the processes at work in the patterning.

\section{Decay at $\Reg$ as a Statistical Physics Problem: Directed Percolation~\label{S4}}

On general grounds, bifurcations in nonlinear dynamics and phase transitions in thermodynamics can be connected via dissipative dynamical systems defined in terms of gradients of a potential which, on one side, govern the most elementary bifurcations and, on the other side, the classical Landau theory.
In Landau's classification of phase transitions~\cite{WPT}, super-criticality maps onto continuous second-order phase transitions, e.g., ferromagnetic, and sub-criticality onto discontinuous first-order phase transitions, e.g., liquid-gas.
The correspondence is strict in the mean-field approximation neglecting microscopic thermal fluctuations.
Taking them into account implies deep corrections.~Second-order transitions then come in with the notion of \emph{universality} linked to a scale-free power-law behavior of correlations at the transition point, introducing sets of critical exponents.
Universality means that physically different systems macroscopically described using observables with identical symmetries behave in the same way at given physical-space dimension.
For a compact self-contained overview of critical behavior and universality in phase transitions consult \S1 of~\cite{Lu04}.
As to first-order transitions, they experience the effect of fluctuations through the nucleation of germs that drive the phase change when they exceed some critical size.

Near equilibrium, thermodynamic systems fulfill micro-reversibility, a property that is lost sufficiently far from equilibrium, in hydrodynamic systems having experienced instabilities, and \emph{a~fortiori} in turbulent flows.
The globally sub-critical transition typical of wall-bounded flows is specific in that it sets a laminar flow stable against small perturbations in competition with a locally highly fluctuating, but statistically well-characterized turbulent regime.
At given $\Rey$ the laminar flow is locally attracting in the dynamical-system sense and is only submitted to extrinsic fluctuations of thermal origin, or due to residual imperfections, that are in themselves unable to drive the flow toward the turbulent state (at least for intermediate values of $\Rey$ in the transitional range).
In statistical physics of far-from-equilibrium systems, this property qualifies an \emph{absorbing} state~\cite{Lu04}, where the word `state' qualifies the system as a whole with a global (thermodynamic) meaning.

\begin{figure}
\centering
\includegraphics[width=0.6\textwidth]{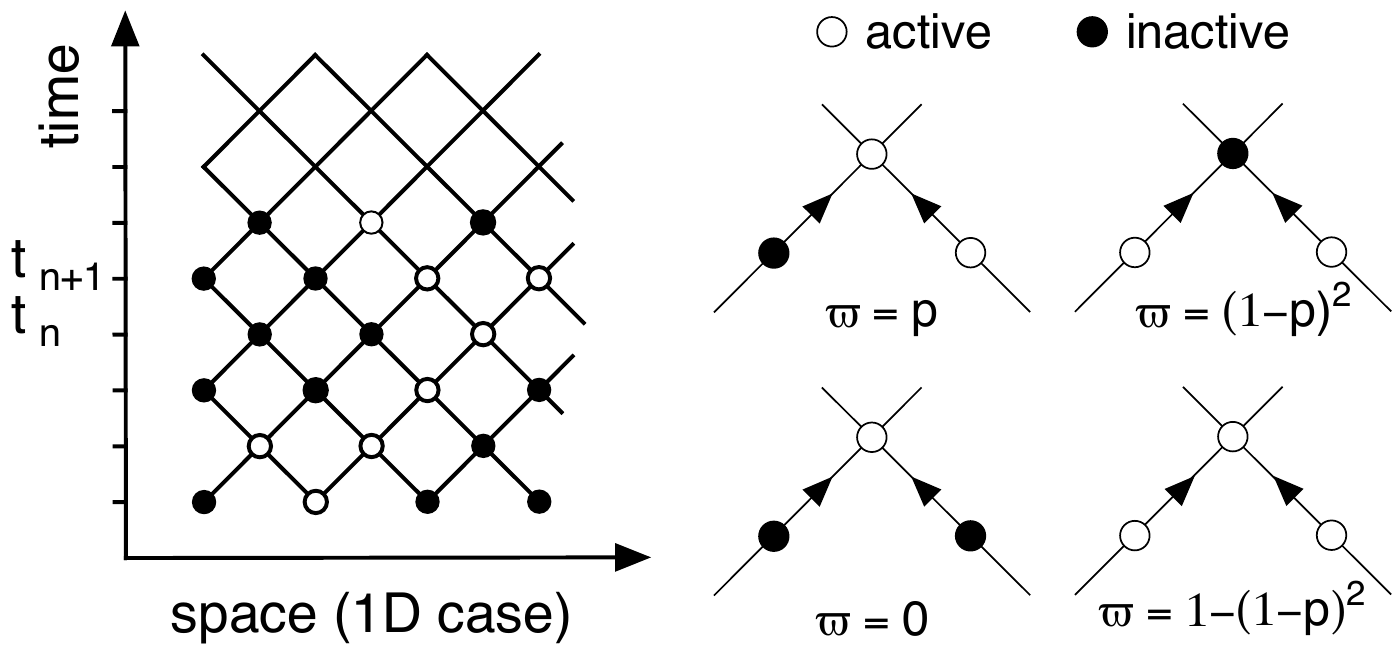}

\vspace{3ex}

\begin{minipage}{0.4\textwidth}
\includegraphics[width=\textwidth]{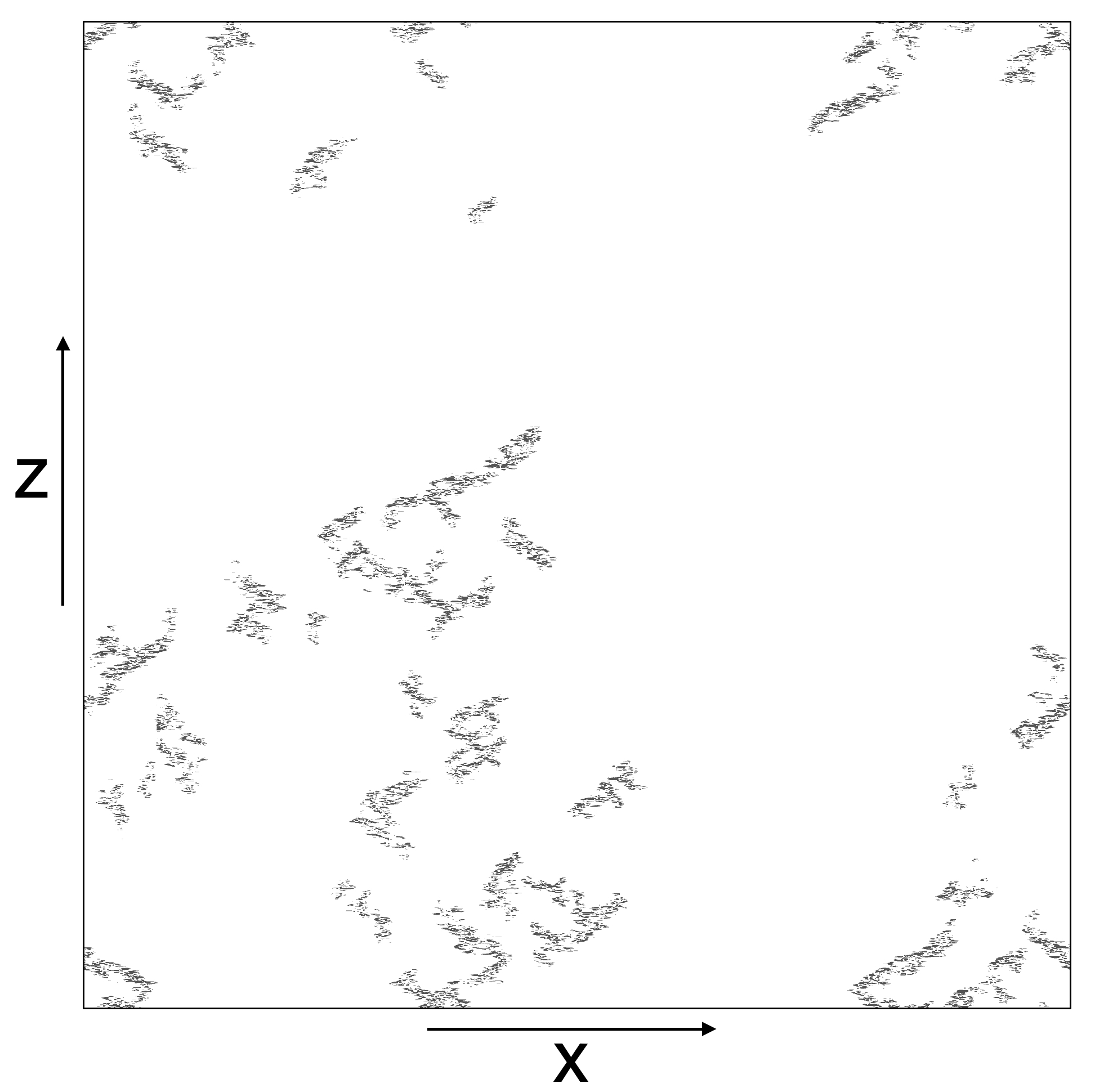}
\end{minipage}\hspace{1em}
\begin{minipage}{0.52\textwidth}
\includegraphics[width=\textwidth]{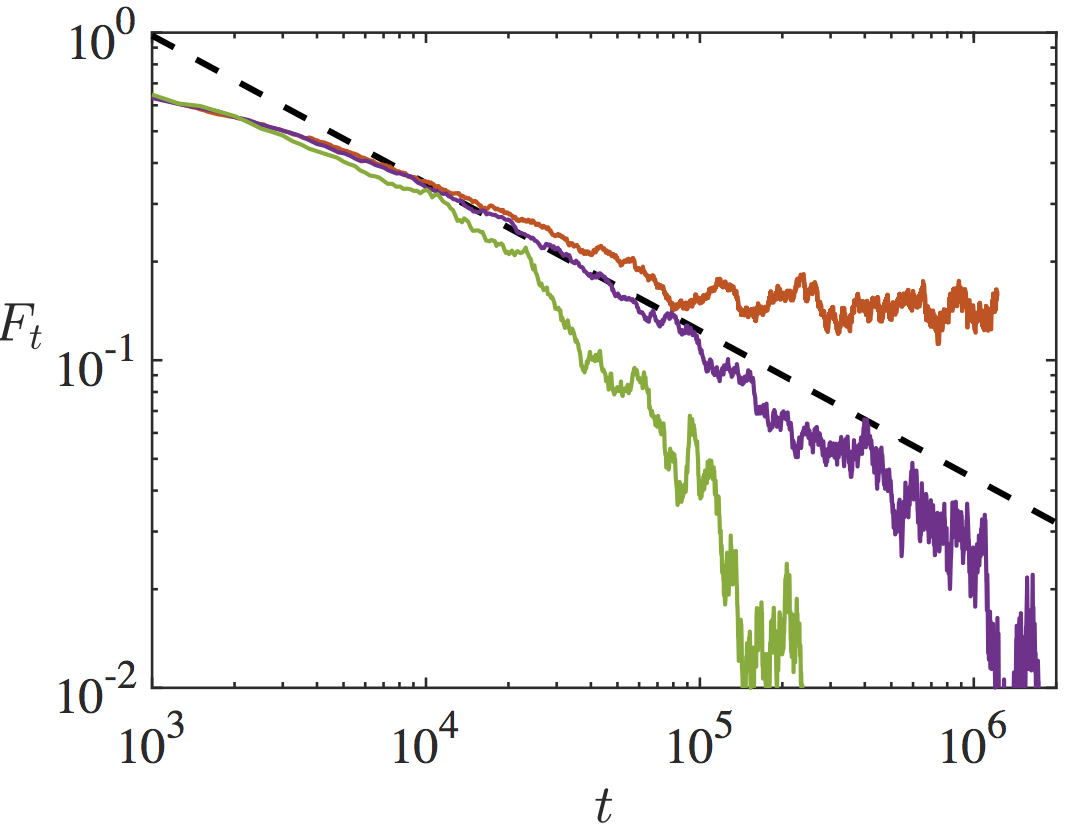}
\end{minipage}
\caption{({\textbf{Top}}) Space-time lattice (\textbf{left}) and contamination rules for different activity configurations as functions of probability $p$ (\textbf{right}) in the 1D case for simplicity.
({\textbf{Bottom}}) Decay of the banded turbulent regime in the stress-free model of PCF by Chantry et al. \cite{Cetalxx2}. (\textbf{Left}) Typical snapshot of streamwise velocity at mid-gap during decay in a $1280h\times1280h$ domain; laminar flow in white. (\textbf{Right}) Variation with time of the turbulent fraction (log-log) during decay; red: saturation somewhat above threshold; green: exponential relaxation to zero somewhat below threshold; purple: near critical, power-law decay followed down until finite departure from threshold is felt, hence late exponentially decreasing tail. The dashed line indicates the theoretical expectation for 2D-DP, here valid over about one decade in time (Bottom panels: courtesy Chantry et al.)\label{f4}}
\end{figure}

On the other hand, the turbulent flow is the seat of large fluctuations of intrinsic origin due to chaos.
Furthermore, all over the coexistence range, this local stochasticity is only \emph{transient}, i.e., can be viewed as a memoryless process with a finite decay probability function of $\Rey$~\cite{Eetal08}. 
Pomeau~\cite{Po86} early suggested that, in view of these characteristics, the whole spatiotemporally intermittent arrangement of laminar-turbulent domains could be interpreted as the result of a purely stochastic process called \emph{directed percolation} (DP) in statistical physics~\cite{Lu04}.~This process can be described in terms of a probabilistic cellular automaton defined on a space-time lattice (Figure~\ref{f4}, top left) of cells that can each be in one of two states, \emph{active} (turbulent) and \emph{inactive} (laminar).
Here `state' has an obvious local meaning, like for spins that can be `up' or `down' (`active/inactive' is also often termed `on/off', even `alive/dead').
A given cell in the inactive state cannot become active by itself, but only by \emph{contamination} with some probability $p$ from one of its neighbors in the active state (Figure~\ref{f4}, top right), as can be the case for trees in forest fires, or individuals in epidemics.
Pomeau went further in conjecturing that the laminar-turbulent transition was in the DP universality class, i.e., had its main properties characterized by the same set of exponents as the abstract statistical-physics process, which triggered numerous studies of analogical models, numerical simulations of NSE, or laboratory experiments.
As to universality, the Janssen--Grassberger conjecture \cite{Lu04} stipulates that all systems with short-range interactions, characterized by a single order parameter, experiencing a continuous transition to a non-degenerated absorbing state, belong to the same class in the absence of additional symmetries or quenched disorder. The main corresponding critical exponents are $\beta$ describing the variation of the turbulent fraction $F_{\rm t} \propto \epsilon^\beta$, where $\epsilon$ is the relative distance to threshold, $\mu_\perp$ and $\mu_\parallel$, accounting for the power-law distribution of absorbing (here, laminar) sequences at threshold, $\mathcal N(\ell) \propto \ell^{-\mu}$ either in space or in time, $\ell$ being the mean size of inactive clusters at a typical time  for $\mu_\perp$, or the mean duration of intermissions (period of inactivity) at a typical location for $\mu_\parallel$. One gets $\beta^{\rm 1D}\approx0.276$, $\mu_\perp^{\rm 1D}\approx1.75$, $\mu_\parallel^{\rm 1D}\approx1.84$, and $\beta^{\rm 2D}\approx0.583$, $\mu_\perp^{\rm 2D}\approx1.20$, $\mu_\parallel^{\rm 2D}\approx1.55$, see \S{}A.3.1 of~\cite{Lu04} for more information.

The study of critical properties of directed percolation comes in two ways, both statistical and requiring a large number of realizations, either (i) by triggering the active state in a germ, a (small set of contiguous) cell(s), or (ii) by observing the decay of a uniformly active state.
The first procedure, which corresponds to the triggering of turbulent spots, has been followed since the early days~\cite{CWP82,TA92,Detal92}, though outside the DP framework.
Numerical approaches devoted to the determination of germs able to drive the transition to sustained turbulence has comparatively received less systematic attention and, in view of reliable statistics, seem much more demanding in terms of analytical shapes to be tested than localized seeds in the 0/1 context of DP.
The search for \emph{edge states} \cite{Setal06}, flow configurations that are sitting on the laminar-turbulent boundary in a phase-space perspective, is a step in that direction \cite{Detal09}.
The second procedure, the decay from a featureless turbulent state, is more easily implemented and has accordingly received more attention in the last few years.

As a first step, analogical models have been considered~\cite{CM94}. 
They were expressed in terms of coupled map lattices where the local map implemented the active/inactive nature of the states at the lattice nodes.
The coupling to neighbors was usually diffusive and 1D or 2D lattices were considered in view of their relevance to 1D pipe flow~\cite{Baxx1,Baxx2}, or to 2D planar flows \cite{BC98} recently re-examined in \S2 of \cite{Cetalxx2}. 
Considering these computationally light cases helps one better figure out the requirements of large aspect ratios and long simulation durations for a proper characterization of the critical behavior at $\Reg$.
These requirements turn out to be extremely demanding, which explains why early PCF experiments were inconclusive as to a 2D-DP critical behavior.
That the DP framework be relevant for PCF was first obtained within the quasi-1D Barkley--Tuckerman framework~\cite{Setal13}, which was quantitatively confirmed soon later \cite{Letal16}.
In addition to the numerical experiment, a quasi-1D CCF configuration was considered with $\eta=0.998$, hence $\Gamma_\theta\approx5000$ and very short cylinders, $\Gamma_z=L/2h=8$, yielding exponents in excellent agreement with the theoretical values for 1D-DP.
\clearpage

In the 2D case, up to now, a single experiment in the PPF case has concluded to the relevance of the DP universality class~\cite{ST16}.
The observation rested on the decay of turbulence produced by a grid at the entrance of a wide ($\Gamma_z\approx180$) and long ($\Gamma_x\approx1180$) channel and a detection of turbulent domains in the part of the channel closest to the exit.
Exponents corresponding to 2D-DP have been found, but the critical point for DP, $\Rey_{\rm c}^{\rm DP}\approx830$, was clearly larger than $\Reg\lesssim660$ above which localized turbulent states are now known to be sustained \cite{TI14,Xetal15, Ketalxx,Petalxx}.
Since by definition $\Reg$ is the threshold below which the base state is unconditionally stable in the long term, this could mean that initial conditions produced in the experiment belonged to the inset of a specific scenario with the flow staying outside the basin of attraction of the localized states mentioned above.
As of today, I am not aware of numerical simulations of 2D-patterned PPF turbulence in systems with aspect ratios sufficiently large to conclude on its decay in a DP perspective, though the use of streamwise periodic boundary conditions should help solving the problem of streamwise advection and the associated experimental aspect ratio limitation (channel length).

Things are different for laminar-turbulent patterns in PCF, at least if one accepts some dose of modeling.
As already mentioned, such a modeling has mainly been developed along two lines, controlled wall-normal under-resolution of the NSE with no-slip boundary conditions~\cite{MR11}, and consideration of stress-free boundary conditions with a subsequent reduction of the number of wall-normal modes \cite{MD01,Cetalxx1}. 
In both cases the computational load is significantly decreased, thus allowing the consideration of larger aspect ratios in order to check the behavior.
Along the first avenue, at reduced wall-normal resolution, the exponent $\beta$ attached to the variation of the turbulent fraction close to the threshold for band decay in a $1000h\times1000h$ domain has recently be found to fit 2D-DP universality by Shimizu~\cite{Setalxx} at a shifted $\Reg$ consistent previous studies~\cite{MR11}.
In the second modeling approach, spectacular results have been obtained by Chantry et al. \cite{Cetalxx2} within the framework of their stress-free reduced model that they simulated in huge domains up to $5120h\times1280h$ and $2560h\times2560h$ (Figure~\ref{f4}, bottom).
They measured all of the exponents of the 2D-DP universality class to extremely good accuracy and obtained excellent data collapse of scaling functions \cite{Lu04} proving their claim.
They also explained why laboratory or numerical experiments in too small domains~\cite{Betal98,Detal10,MFDR} could erroneously suggest a discontinuous transition as sketched in Figure~\ref{f1} (b1-b2).
However, they also documented further that, within coupled-map-lattice modeling~\cite{CM94,BC98}, the DP universality class is particularly fragile in 2D and thus prone to break down as a discontinuous transition (\S2 of \cite{Cetalxx2}). 
Since modeling specificities, such as the bad account (under-resolution) or neglect (stress-free) of boundary layers close to the walls, could affect the properties of the transition, simulations of the realistic case with no-slip conditions at full resolution are underway~\cite{Setalxx}.

\section{Emergence of Patterns from the Featureless Regime\label{S5}}

Pattern formation is a standard problem in non-equilibrium dynamics \cite{Ma10,CG09}.
Usually, e.g., in convection, the state of the considered system lies on the thermodynamic branch where the effects of noise, of thermal origin, are small and bifurcations away from this state are essentially governed by deterministic dynamics.
On general grounds, one expects a super-critical bifurcation governed by an ordinary differential equation, viz. a \emph{Landau} equation: 
$\frac{{\rm d}A}{{\rm d}t}=r A - A^3$ where $r$ is a reduced control parameter. $A(t)$, the amplitude of the deviation from the basic state, is a function of time~$t$.
When a periodic pattern forms, amplitude $A$ is the intensity of corresponding Fourier mode, e.g., convection rolls with wavelength $\lambda_{\rm c}$.
In large aspect-ratio systems, the spatial coherence induced by the local instability mechanism cannot be maintained by lateral boundary effects.
The intensity of the developing structure get modulated with an expected tendency to relax toward the arrangement favored by the instability mechanism in a diffusive fashion.
Typically $A$ becomes a function of time \emph{and} space, $A(x,t)$, governed by a partial differential equation of \emph{Ginzburg--Landau} (GL) type: $\partial_t A=r A + \Delta A - A^3$, in which $\Delta$ is a Laplacian operator accounting for diffusion, in 1D or 2D depending on the geometry.
This simplified description can be extended to deal with competing modes and associated amplitudes with specific symmetries.
Such formulations can (at least in principle) be derived from the NSE via multi-scale expansions resting on scale separation, i.e. $\lambda_{\rm c} \ll $ modulation scales, as discussed e.g. in Chapter~6 of \cite{Ma10}.
Weak extrinsic noise can be introduced as an additive perturbation.

The high degree of generality of this approach \cite{CG09} gives strong motivation to its use at a \emph{phenomenological level} when strict applicability conditions are not fulfilled,
here in an overall globally-sub-critical context for an apparently continuous bifurcation which is super-critical-like, but at decreasing control parameter \emph{and} from a uniform turbulent background.
 Prigent et al.~\cite{Petal03,PD05} introduced such a description of patterning in CCF, directly stemming from their observations with $\eta=0.983$, as summarized at the end of Section \ref{S2}.
Introducing a set of two coupled GL equations for two amplitudes, one for each orientation, and adding a noise term to account for the intrinsic stochasticity in the turbulent background, they were able to fit all of the phenomenological coefficients introduced in their expressions against the experiments, including the effective noise intensity, and to account for the whole variation of the pattern's amplitude with a reduced control parameter $\propto \Ret-\Rey$.
The fits used the amplitude of the dominant Fourier modes of the turbulence intensity in a plane containing the cylinders' axes, with demodulation to separate the two possible helical pattern components.
They showed that the amplitude followed the square-root behavior expected from GL theory, extrapolating to an apparent threshold $\Ret$ beyond the values of $\Rey$ where the pattern becomes visible by eye. 
This observation was understood as an effect of high-level noise from the background turbulence implying strong orientation fluctuations and a subsequent reduction of the pattern's amplitude.
When transposed to PCF, this provides an explanation to the difference between the value $\Ret\approx 440$ obtained in the Barkley--Tuckerman oblique domain \cite{Tetal09}, and the value found consistently in the range $405$--$415$ in experiments~\cite{Petal03,PD05} or numerical simulations in streamwise-spanwise extended domains~\cite{Detal10,Ma15} that keep full track of orientation fluctuations.

\begin{figure}
\centering
\includegraphics[width=0.45\textwidth]{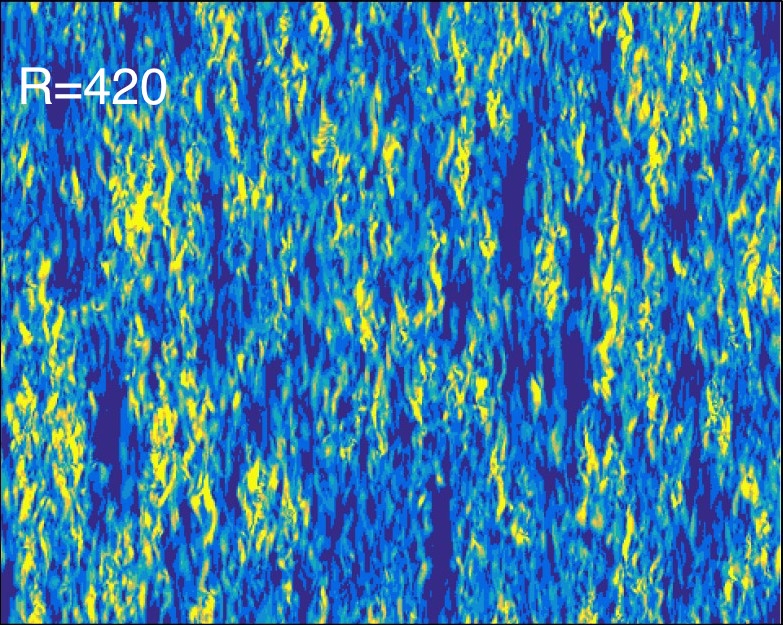}\hspace{1em}
\includegraphics[width=0.45\textwidth]{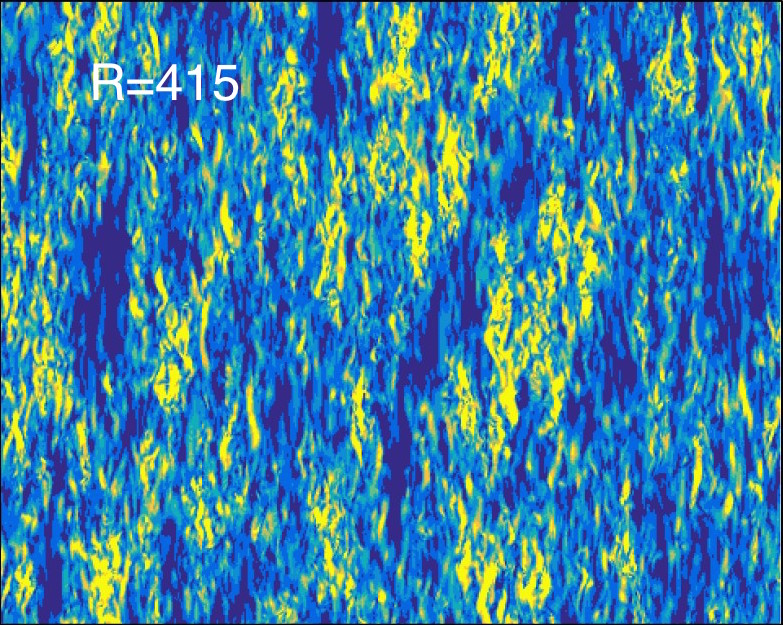}\\[2ex]
\includegraphics[width=0.45\textwidth]{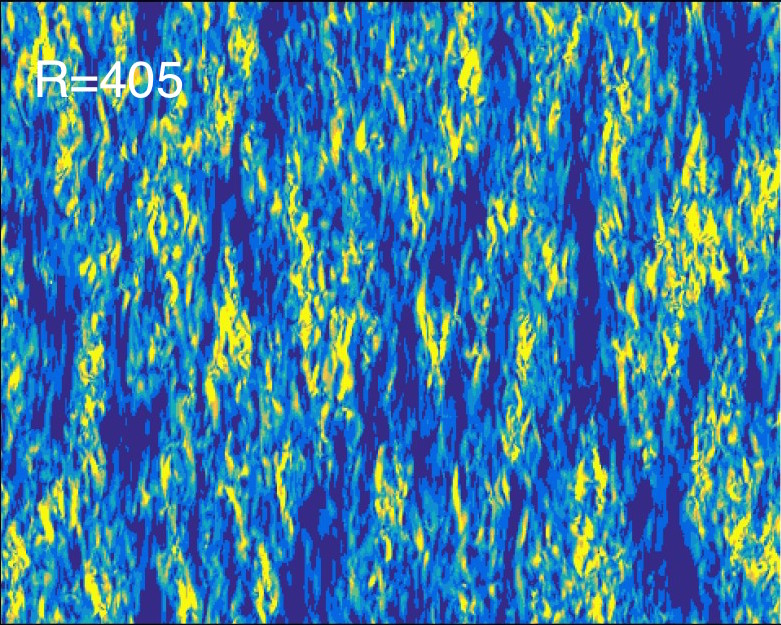}\hspace{1em}
\includegraphics[width=0.45\textwidth]{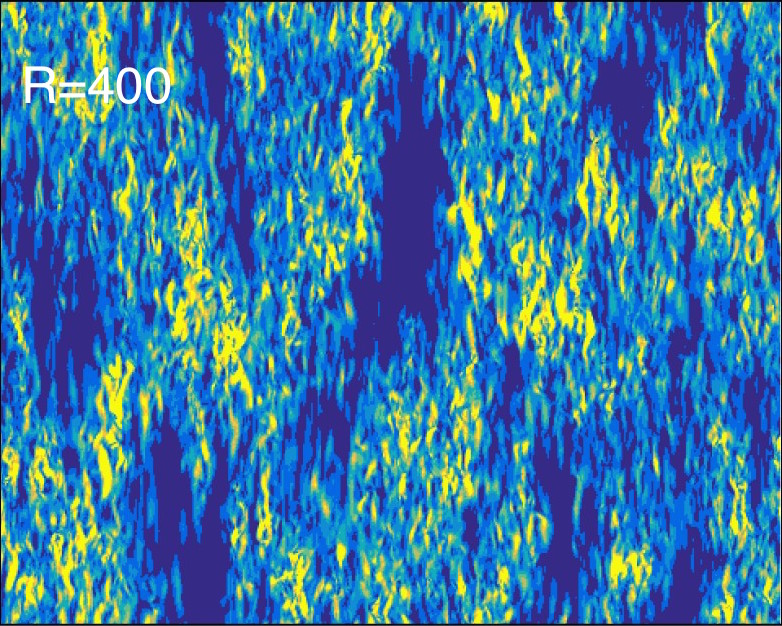}\\[2ex]
\includegraphics[width=0.5\textwidth]{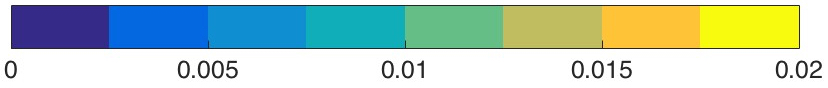}

\caption{Emergence of the pattern at $\Ret$, as seen from color levels of the cross-flow kinetic energy averaged over the gap in well-resolved numerical simulations of NSE \cite{Ma15}. 
The streamwise direction is vertical and the color scale for the local transverse kinetic energy averaged over the gap is identical for all pictures. From  top to bottom and left to right:
$\Rey=420$, only short-lived, mostly streamwise-aligned, elongated laminar troughs (deep blue).
$\Rey=415$, some laminar troughs become wider, others get inclined.
$\Rey=405$, laminar troughs transiently form alleys of both orientations.
$\Rey=400$, turbulent fraction decreases significantly owing to bulkier laminar troughs.
The illustrations shown are typical of flow patterns around $\Ret$. \cite{Petal03} 
\label{f5}}
\end{figure}

Better understanding fluctuations around $\Ret$ should give insight in the nature of the transition to turbulence and the mechanisms presiding the emergence of a pattern.
To this aim, I performed numerical simulations of PCF in domains of size $\sim128h\times160h$ hosting two to three turbulent bands~(App.A of \cite{Ma15}).
Figure \ref{f5} depicts typical snapshots from my simulations down to $\Ret$ that could however not be precisely located due to size effects and lack of statistics to deal with the high level of fluctuations. 
As a matter of fact, intermittent elongated laminar patches appear well above $\Ret$ in what remains predominantly featureless turbulence.
They are small, short-lived, and mainly streamwise without any apparent ordering above $\Rey\approx420$.
For $\Rey\lesssim415$, though still intermittent, they become bigger, occasionally oblique, with longer lifetime, and show a tendency to cluster, either with the same orientation or forming chevrons when the orientations were different.
As $\Rey$ is further lowered, the turbulent fraction decreases due to the widening of the laminar domains that seem to progressively percolate (in the ordinary, not directed, sense) through the remaining turbulent flow. 
Fourier characterization of this modulated turbulence warrants further scrutiny in view of a comparison with results from experiments in CCF \cite{Petal03,PD05} and the quasi-1D numerical approach \cite{Tetal09}.

The origin of localized, short-lived, MFU-sized laminar patches can be traced back to the behavior of chaotic solutions to the NSE at the MFU scale, with their irregular alternations of bursting and low-activity excursions \cite{Ketal12}, but the formation of more extended laminar domains remains to be elucidated.
The nature of the flow inside laminar patches of intermediate size is of special interest since, on general grounds, it can be analyzed as a superposition of the laminar base flow and a large scale correction (the faint yellow lines on the blue background in Figure~\ref{f3}).
In the lower part of the transitional range $[\Reg,\Ret]$ where the pattern is well established, the large-scale flow is easily extracted by time-averaging \cite{BTxx2} or by filtering out small scales around growing spots \cite{DS13,CM15}.
In systems bounded by plates in the wall-normal direction, PCF, PPF, etc., a non-vanishing 2D divergence-free component can be isolated out of the large-scale flow by averaging over that direction \cite{DS13}.
This component might be crucial for the organization of laminar patches into regularly arranged bands since patterns disappear when large-scale flows are not constrained to stay 2D, but allowed to escape in the third direction as for ASBL \cite{Ketal16a} or too weakly stratified Ekman layer \cite{Detal14}.
Unfortunately, in the upper transitional range around $\Ret$ this large-scale flow component is difficult to educe without a combination of averaging and filtering free from arbitrariness when laminar depressions are still small.
If detected unambiguously, it could however serve to characterize the transition at $\Ret$ in much the same way as it was used to identify the transition from 1D to 2D in annular Poiseuille flow when curvature is decreased~\cite{Ietalx}.
How much does it contribute to the percolation of laminar patches into bands at $\Ret$ and the details of the laminar-turbulent organization, wavelength and angle, whether it is a cause or a consequence, are questions that directly leads to the discussion below.

\section{Understanding Laminar-Turbulent Patterning:\\ Theoretical Issues and Modeling Perspectives\label{S6}}

Ideally, the physical explanation for patterning should derive from the NSE.
Unfortunately this does not seem easy since, whereas the analytical solution for laminar flow can be straightforwardly obtained, neither locally turbulent flow nor the flow in the interface region can be obtained without some empirical averaging or modeling.
Closure models that work in engineering conditions where turbulence is rather homogeneous and sufficiently developed do not provide appropriate solutions in the transitional regime where coexistence is the rule.
This was shown in~\cite{Tetal09} where a standard $K$--$\Omega$ approach was used to treat the laminar-turbulent mixture without producing any modulation of the turbulence intensity.
Other approaches must therefore be followed.

Working by analogy is the way followed by Barkley to obtain his model for HPF~\cite{Baxx1,Baxx2}.
First identifying the similarity between puffs in a pipe (HPF) and nerve impulse propagation, he treated the pipe as a 1D reaction-diffusion (RD) system (Chapter 9--14 of \cite{Mu93}).
In the \emph{excitable} regime, the dynamics only produces localized concentration perturbations, pulses, followed by refractory stages during which the medium can recover the properties necessary for pulse propagation.
Upon changing the reaction and diffusion rates, the RD system can enter a \emph{bi-stable} regime with locally homogeneous domains of reactants or reaction products generically separated by propagating fronts, e.g., flame fronts, the domain filled with the most stable state, e.g., burnt gases, invading that of the least stable one, i.e., fresh gases.
Barkley \cite{Baxx1} introduced just two variables, one `$q$' measuring the local turbulence level and another one `$u$' characterizing the mean shear, coupled by two partial differential equations functions of the streamwise coordinate $x$ and time $t$. 
An appropriate choice of the intrinsic dynamics for $q$ permitted the control of the \emph{excitable vs. bistable} behavior by means of a single parameter playing the role of $\Rey$.
A coupling of $u$ and $q$ mimicked empirical knowledge on the relaxation of $u$.
This led Barkley to the sought-after model \cite{Baxx1}, allowing him to reproduce the overall behavior of HPF, from puff dynamics (excitable regime) to the transformation into slugs (bistable regime).
The adjustment of a few coefficients allowed the quantitative reproduction of the different dynamical regimes observed and the model was simple enough to permit a detailed analytical treatment that really explains the behavior of HPF all along its transitional range \cite{Baxx2}.

A support of the reduced description sketched above from the NSE would however be welcome.
In that form, it can also not help us understand the 2D patterning typical of planar flows.
As a first step in that direction, I proposed~\cite{Ma12} to keep the RD framework, but to exploit another of its features: the possibility of a Turing instability ending in a pattern when diffusivities of the reactants are of different orders of magnitude (Chapter~14 in \cite{Mu93}).
To be more specific, I chose to describe the local reaction using Waleffe's implementation of the SSP~\cite{Wa97}. From its four variables, I enslaved two of them to the mean flow correction $m$ analogous to Barkley's $u$ and to a second variable $w$ analogous to $q$.
Furthermore, I introduced diffusion along a fictitious space coordinate, slow for $m$ and fast for $w$ as guessed from the physical nature of these variables and, magically, patterning emerged.
A beneficial aspect of my approach was that the nonlinear reaction part could be traced back to NSE by following Waleffe, but, while alleging that patterns might result from the interplay of effective diffusion and nonlinear reaction, by construction it could predict neither the wavelength nor the orientation of the so-obtained pattern.
In an attempt to fix the wavelength in a 1D periodic domain, Hayot and Pomeau~\cite{HP94} introduced a phenomenological feedback from large-scale secondary flows in a sub-critical GL formulation, $\partial_t A = r A + \partial_{xx} A + (A^2-B) A - A^5$, where $B=L^{-1} \int_0^L A^2 {\rm d} x$ expressed the pressure loss through Reynolds stresses in the turbulent fraction of the whole domain.
While a variation of the turbulent fraction with $r$ could indeed be predicted, a single laminar-turbulent alternation was obtained, hence no nontrivial wavelength.
Realistic, though simple enough, modeling leading to an explanation of pattern formation could hopefully come from a combination of the ingredients mentioned up to now, especially if one could make them stem from the NSE in some way.

A promising strategy follows from the remark that under-resolved numerical simulations of NSE give already precious qualitative information on the processes at work \cite{MFDR}, and even quantitative results \cite{Setalxx} once the systematic downward shift of $\Reg$ and $\Ret$ \cite{MR11} is taken into account.
The most common simulations are based on spectral representations of the wall-normal dependence.
The robustness of patterning against under-resolution stems from the fact that the dynamics seems controlled rather by the behavior of the `interior flow' \cite{Cetalxx1} than by what happens close to the solid boundaries.
 Further, owing to the intermediate values of $\Rey$ involved, neither too small nor too large, this behavior can be accounted for by the very first modes of the functional expansion of wall-normal space dependence that can be dealt with analytically rather than in the black-box fashion of a simulation software.
This treatment is all the more feasible that the basis functions can be chosen for analytical simplicity rather than for computational efficiency.
Trigonometric lines are appropriate to stress-free boundary conditions, as proposed by Rayleigh \cite{RaXX3} and exploited in \cite{Wa97,MD01,Cetalxx1}.
If no-slip boundary conditions are judged more realistic, simple polynomials \cite{LM071,LM072,SM15} can be preferred to Chebyshev polynomials generally used in~simulations.

The standard Galerkin approximation procedure eliminates the explicit dependence on the wall-normal coordinate and replaces the 3D velocity and pressure fields by 2D mode amplitudes in the planar case \cite{SM15,Cetalxx1} and even 1D for pipe flow \cite{Cetalxx1}.
Simultaneously the NSE is replaced by a set of partial differential equations with reduced spatial dimensionality.
Importantly, truncation of the expansion, retaining just a small number of amplitudes and equations, implements the dominant features of the dynamics, namely the SSP, while preserving the general structure of the NSE, notably its symmetries relevant to the case at hand and kinetic energy conservation by the advection term, as discussed in \S3.2 of \cite{MD01}.
Numerical simulations of so-obtained models show that realistic patterning in PCF is obtained by keeping just seven amplitudes \cite{Cetalxx1,SM15}, while the three lowest ones account locally for Waleffe's implementation of the SSP \cite{MD01}, and globally for large scale flows around a turbulent patch~\cite{LM072}.

The analytical approach in \cite{SM15} emphasizes the generic character of the such low order models where a particular coefficient set relates to a given system (flow geometry and boundary conditions, e.g., stress-free {vs.} no-slip, cf. Table~1 in \cite{LM071}), which can be tentatively changed to test the effect of specific coupling terms.
Beyond plain simulations, the formulation can be the starting point for further modeling in view of building RD-like simplified models giving some foundation to the nonlinear interaction terms introduced on semi-empirical grounds for HPF \cite{Baxx1} or from purely phenomenological arguments for PCF \cite{Petal03,PD05}.
This derivation should focus on slow and large scale properties relevant to patterning, therefore eliminating all spatiotemporally fast interaction terms at the MFU scale, as partially done in \cite{LM072}, or in \cite{HP94,Ma12}.
The problem lies in a realistic modeling of Reynolds stresses generated by the SSP as a local source driving the large scale flow, which might be achieved by completing Waleffe's local approach \cite{Wa97} with a closure assumption expressing the feedback of large scale flows on the turbulence level at MFU scales.
Including such physical insight would help one skirt around the limitations of closures usually referred to in turbulence modeling, such as the $K$--$\Omega$ scheme used in \cite{Tetal09}.

The interest of such a modeling would not primarily be in view of appreciating/questioning universality at $\Reg$ when the pattern decays since, in very large systems, the turbulent fraction is so low that long range interactions associated to large scale flows are expected to be extremely weak (Figure~\ref{f4}, bottom-left) and are not really suspected to violate the terms of the Janssen--Grassberger conjecture mentioned in Section~\ref{S4}.
On another hand, offering a reliable representation of the dynamics at scales somewhat larger than the MFU, it would provide indications about the dependence on $\Rey$ of probabilities for turbulent patches to grow, recede, or branch, for turbulent bands to break and recover from laminar gaps, etc. \cite{MFDR}.
Besides giving a microscopic foundation to the macroscopic behavior at $\Reg$, its main interest would certainly be to answer the question of patterning emergence from the FT regime when $\Rey$ decreases from large values.
In this respect, the goal would be to eliminate all irrelevant information and derive an effective GL formulation valid all along a large part of the transitional range, accounting for laminar-turbulent alternation with possible superposition of different orientations around $\Ret$, for the selection of a given orientation somewhat below $\Ret$, and for wavelength and orientation changes as $\Rey$ decreases, since all of this can be contained in the coefficients of the effective GL model \cite{Petal03,PD05}.

Another question is why does patterning occurs at all, and whether it achieves some sort of dynamical optimum (minimization of an effective potential with thermodynamic flavor), as would stem from the weakly nonlinear GL formalism with added noise \cite{Petal03,GT90}.
It is however not clear how to apply this approach when working at decreasing $\Rey$ from a turbulent state.
In previous studies~\cite{GT90} the bifurcating and bifurcated states were affected in the same way by weak additive noise.
For example in convection, the bifurcation is super-critical and, near threshold, the rest state and the B\'enard cells remain qualitatively and quantitatively close to each other in a phase-space perspective and are perturbed by small extrinsic imperfections and intrinsic low-amplitude thermal noise in the same way.
In wall-bounded flows the branch of nontrivial states is qualitatively always far from the laminar flow branch in phase space, even when the distance is quantitatively evaluated as a vanishingly small turbulence fraction immediately above $\Reg$  as sketched in Figure~\ref{f1} (b2).
Meanwhile, far above $\Ret$ the FT regime displays large, space-time localized, intrinsic fluctuations that make the flow `remember' the presence of the laminar (absorbing) regime, far from a situation where detailed balance would hold.
Any thermodynamic viewpoint about transitions consequently remains a challenge, at least compared to spin systems or other microscopic systems at equilibrium.

Beyond these formal considerations, I would like to conclude by first stressing that organized patterning is a common feature of transitional wall-bounded flows, with laminar-turbulent coexistence holding both in physical space and, usually, over some finite range of Reynolds numbers.
Next global sub-criticality is linked to the absence of any relevant instability against infinitesimal perturbations to the laminar base flow in the whole series of systems that I have considered, PCF being just a paradigmatic case.
In addition, the universal behavior of turbulence decay at $\Reg$, much debated for a long time since Pomeau's early conjecture \cite{Po86}, is on the verge of being demonstrated.
However, though conceptually satisfactory, this property concerns a narrow vicinity of $\Reg$ and appears to be much less important than the nature of physical processes involved at intermediate $\Rey$ and large scales (i.e., $\gg$ MFU), at the laminar-turbulent interface dynamics, especially in spot growth and pattern formation.
Further study of these subjects, experimental, numerical, or theoretical {\it via\/} simplified, but realistic modeling, seem particularly necessary in view of controlling the transition in less academic cases, a matter of great practical interest for applications.\\[1ex]
\noindent {\bf Acknowledgments}\\
I should first thank Y.~Pomeau (ENS, Paris, France) who gave me the initial impetus in all of the aspects of the problem considered in this review. Former collaborators, H.~Chat\'e, F.~Daviaud and his group (CEA-Saclay, Gif-sur-Yvette, France), D.~Barkley (University of Warwick, Coventry, UK) and L.S.~Tuckerman (ESPCI, Paris, France), as well as the participants in the JSPS-CNRS bilateral exchange collaboration {\sc TransTurb}, G. Kawahara and M. Shimizu (Osaka University, Osaka, Japan), T. Tsukahara (Tokyo University of Science, Tokyo, Japan), and Y. Duguet (LIMSI, Orsay, France), R. Monchaux and M. Couliou (ENSTA, Palaiseau, France) also warrant deep acknowledgments for their contribution to my present understanding of this research field.\\[1ex]
\noindent {\bf Abbreviations}\\[2ex]
\noindent 
\begin{tabular}{@{}ll}
1D/2D/3D & One/two/three-dimensional (depending on 1/2/3 \emph{space} coordinates)\\
ASBL & Asymptotic suction boundary layer (along porous wall with through flow)\\ 
CCF & Cylindrical Couette flow (flow between differentially rotated coaxial cylinders)\\
CPF & Couette--Poiseuille flow (flow between moving walls under pressure gradient)\\
DP & Directed percolation (stochastic competition between decay and contamination)\\ 
FT & Featureless turbulence (uniformly turbulent flow)\\
GL & Ginzburg--Landau (formulation accounting for modulated periodic patterns)\\
HPF & Hagen Poiseuille flow (flow in straight cylindrical pipe under pressure gradient)\\
MFU & Minimal flow unit (domain size below which no sustained nontrivial flow exist)\\
NSE & Navier--Stokes equation, primitive equation governing the flow behavior\\
PCF & Plane Couette flow (shear flow between counter-translating walls)\\
PPF & Plane Poiseuille flow (flow between plane walls under pressure gradient)\\
RD & Reaction-diffusion system (field description of reacting chemical mixtures)\\
SSP & Self-sustainment process, mechanism for nontrivial nonlinear states
\end{tabular}

\end{document}